\documentclass[epjST]{svjour}
\usepackage{graphics}
\usepackage{amssymb,amsmath}
\usepackage{wrapfig}
\begin{document}
\title{A brief introduction to the model microswimmer {\it Chlamydomonas reinhardtii}}
\author{Rapha\"el Jeanneret \and Matteo Contino \and Marco Polin\thanks{\email{mpolin@warwick.ac.uk}}}
\institute{Physics Department, University of Warwick, Gibbet Hill Road, Coventry CV4 7AL, United Kingdom}
\abstract{
The unicellular biflagellate green alga {\it Chlamydomonas reinhardtii} has been an important model system in biology for decades, and in recent years it has started to attract growing attention also within the biophysics community. Here we provide a concise review of some of the aspects of {\it Chlamydomonas} biology and biophysics most immediately relevant to physicists that might be interested in starting to work with this versatile microorganism.    
} 
\maketitle
\section{Meet {\it Chlamydomonas}}
\label{sec:1}

\begin{figure}[t]
\centering
\resizebox{0.45\columnwidth}{!}{
    \includegraphics{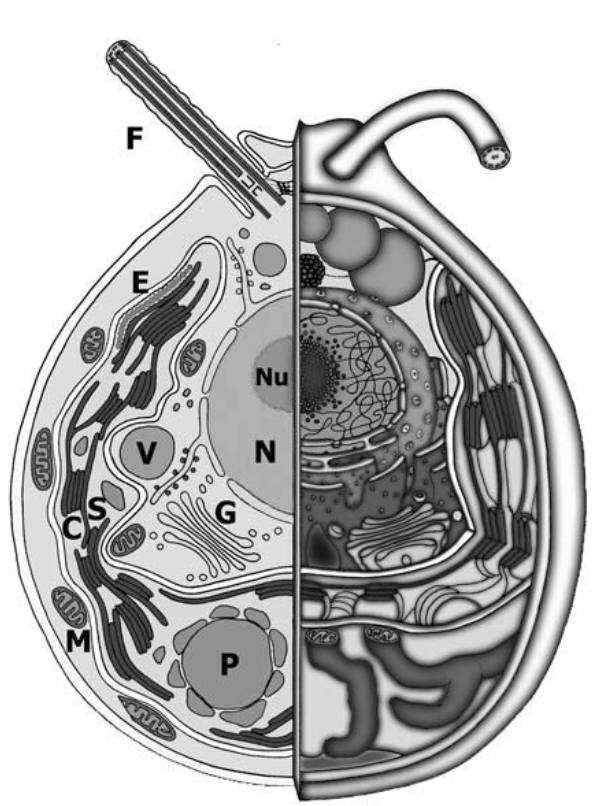}}
  \caption{Schematic layout of a vegetative {\it Chlamydomonas} cell. $F$: flagella; $E$: endoplasmic reticulum; $V$: vacuole; $S$: starch granule; $C$: chloroplast; $M$: mitochondria; $P$: pyrenoid; $G$: Golgi apparatus; $N$: nucleus; $Nu$: nucleolus. (From \cite{harris09})}
  \label{fig1}
\end{figure}

{\it Chlamydomonas reinhardtii} (order {\it Volvocales},  family {\it Chlamydomonadaceae}) is a unicellular green alga which has emerged in the last 60 years as a premier model system within a large variety of areas in molecular and cell biology, including structure and function of eukaryotic flagella, biology of basal bodies/microtubule organising centres, organelle biogenesis \cite{chan10,marshall11,gohering12}, photosynthesis \cite{allorent13,allahverdiyeva14}, cell cycle control \cite{dodd13,niwa13}, cell-cell recognition \cite{snell85}. 
The complete genome has been sequenced relatively recently \cite{merchant07}. There are arguably at least three main reasons leading to this development: i) {\it Chlamydomonas} (CR) is easy to grow in the lab (see further in this section); ii) its cell cycle can easily be entrained to the day/night cycle of the diurnal chambers where it is usually grown, offering a very straightforward way to generate macroscopic suspensions of cells whose progression through the cell cycle is (essentially) perfectly synchronised, thus facilitating a lot e.g. proteomic and metabolomic research; iii) it has proven relatively easy to isolate and characterise mutants, so much so that hundreds of different mutants can be quite simply ordered from algal collections around the world (more on this below). 

All this has  ushered, in the last $\sim10$ years, a new interest in CR on the biophysics front -although ``visionary pioneers'' were working on it already over 20 years ago \cite{pedley92}-. Physicists, mathematicians and engineers have engaged primarily with two areas: photosynthesis and motility. These notes are intended to be a brief introduction to some aspects of the latter. It is not intended to be exhaustive, and it will not be, but hopefully it will provide a starting point for further reading. An excellent reference text is ``{\it The Chlamydomonas Sourcebook}'' \cite{harris09}, a detailed 3 volumes review on the state-of-the-art knowledge on {\it Chlamydomonas reinhardtii}. Although quite technical at times, Vol.~1 provides a very comprehensive review of the main aspects of CR biology \footnote{Vol.~2 reviews CR biochemistry, while Vol.~3 focusses on motility and behaviour.}. A slightly older, but very condensed review on CR can be found here \cite{harris01}. Also very interesting, although not exclusively focussed on CR, is a recent review on volvocine algae in biological fluid dynamics \cite{goldstein15}.

The species {\it Chlamydomonas reinhardtii} was first described by Dangeard in 1888, who named it in honour of the Ukranian botanist Ludwig Reinhardt. There are currently three principal strains used for research: the Sager line; the Cambridge line; and the Ebersold/Levine 137c line. The main wild type strains used in the literature, 21 gr, (UTEX 89, UTEX90), (CC124, CC125), come from each of these three lines respectively. They are all supposedly descendant from a single mating pair ({\it plus} and {\it minus}, akin to male and female) derived from the third (c) zygospore in isolate 137 collected by GM Smith in 1945 from a puddle in a potato field near Amherst, Massachussetts (CR is a soil alga!). 
As such, the different lines should all be the same, but they are not. In particular, the Ebersold/Levine line is well known to contain two mutations ({\it nit1} and {\it nit2}) which prevent the cells from using nitrate as the only N-source. This should be remembered when comparing results between different strains, which might come from different lines and hence have slightly different ``backgrounds'' (the common genetic blueprint of each line). For more informations see \cite{harris09} Vol.~1, pg.~12.

Fig.~1 illustrates the basic cellular architecture in CR. The cell body is approximately a $10~\mu$m diameter spheroid, containing all the standard eukaryotic organelles (nucleus/nucleolus, mitochondria, rough and smooth endoplasmic reticulum, Golgi apparatus etc.). The basal 2/3 of the cell are occupied by a single cup-shaped chloroplast, where light capture and photosynthesis happen. The chloroplast contains a single pyrenoid located towards the base of the cell, where CO$_2$ is fixed, and most of the starch accumulates. This front-back asymmetric architecture causes the centre of mass to be displaced towards the bottom of the cell (bottom-heavy) which induces a slight upward bias in the cells' swimming, through a so-called gravitactic torque \cite{pedley92,durham11}.
Towards the cellular equator we find the eyespot, a rudimentary light-sensitive organelle that the cell uses to perform phototaxis (motion towards/away from light). The eyespot is composed of two main parts. One is a specialised region of the plasma membrane containing (many copies of) channelrhodopsin, a light-gated ion channel protein with good sensitivity in the $450-700\,$nm spectral range. The other is the stigma, a specialised region of the chloroplast containing several stacks of carotenoid-rich granules acting as a dielectric mirror \cite{foster80}. This mirror has a dual role: it concentrates the light on the rhodopsins when the eyespot is facing the light source; and it screens the rhodopsins when the eyespot is facing away from the light source. This results in a $\sim 80$-fold increase in the light signal detected by the cell and hence a more accurate motile response to light \cite{roberts01}. The carotenoids also give the eyespot its characteristic bright orange-red colour. 
Towards the cell apex we find two contractile vacuoles. These organelles are common in freshwater protists, including soil-dwelling species like CR, where they regulate intracellular pressure by periodically ejecting excess water that entered the cell by osmosis \cite{harris09}. In CR they swell (diastole) reaching $\sim2\,\mu$m diameter, and quickly contract (systole, $\sim0.2\,$s) with a period of $10-15\,$s. The precise mechanism leading to water ejection is unclear \cite{kosmic14}. Close to the contractile vacuoles are two basal bodies, from which the two flagella of CR originate. Basal bodies have a cylindrical shape and are composed of 9 microtubule triplets. They not only act as flagellar bases, but during cell division double up as centrioles. As such they are essential organelles. The two basal bodies are directly connected by the distal striated fibre, containing the contractile protein centrin. Additional fibres (rhizoplast), also centrin-based, connect basal bodies to the nucleus. There is evidence that centrin-based fibres can contract {\it in vivo} in response to changes of Ca$^{2+}$ concentration in the cell \cite{salisbury87}. Given the role that distal striated fibres seem to play in flagellar coordination within a single {\it Chlamydomonas} cell \cite{quaranta16,wan16} (see sec.\ref{sec:4}), it is possible that changing their tension might have an impact on flagellar dynamics and synchronisation. Additional sets of fibres connect the basal bodies to four microtubule rootlets, which extend deep within the cell body and are responsible for a precise and reproducible arrangement of cellular organelles (e.g. the correct orientation of the eyespot relative to the flagellar plane, which is essential for phototaxis) (see also \cite{harris09}, Vol.~3, Ch.~2). CR cell body is enclosed in a $\sim200\,$nm-thick cell-wall composed of 7 distinct layers consisting primarily of glycoproteins, with no trace of cellulose. The ultrastructure of the wall is well characterised, but its synthesis and assembly is not understood as well. From each basal body a single flagellum extends outwards for $\sim10-12\,\mu$m. The flagella are motile and usually beat in a characteristic breaststroke fashion at $\sim50\,$Hz. These will be described in more detail in the next section.
Vegetative {\it Chlamydomonas} cells are haploid (i.e. the nucleus contains the same number of chromosomes than their gametes \footnote{CR gametes are also monoploid, i.e. they have only one homolog of each chromosome. This is the same in humans, but not e.g. in wheat, where gametes have three homologs of each chromosome. Remember that human somatic cells have twice the number of chromosomes found in their gametes, i.e. they have a so-called diplontic life-cycle. Since human gametes are monoploid, human somatic cells are diploid.}), and can reproduce indefinitely in this state. This asexual reproduction has a cycle of approximately one day, and can be entrained to be exactly one day if the cells are grown within a diurnal chamber set to a day/night cycle of 24hr. This is (necessarily) a clonal reproduction, whereby the mother cell undergoes $n$ subsequent cell divisions during the night, producing $2^n$ daughter cells. These then hatch from the mother and the cycle repeats. The number of divisions $n$ depends on the cellular volume reached by the mother cell as it commits to cell division (size checkpoints). In our hands we find it to be at most 3 (i.e. 8 daughter cells). This fascinating process, including the control possibilities afforded by light, has been studied extensively (see also \cite{harris09}, Vol.~1, Ch.~2). 
CR can also undergo sexual reproduction, whereby a pair of cells of opposite mating types ({\it plus} and {\it minus}) fuse together to form a temporary quadriflagellate zygote (diploid) which remains motile for $\sim2\,$hr and then forms a zygospore with a tough external wall. In this state the cell has available proteins from each of the two original haploid cells, and thus mutations carried by only one of these can be recovered in the zygote by the proteins of the other. This is called dikaryon rescue, and it has been used extensively to study allelic dominance especially in flagellar mutations (see \cite{harris09}, Vol.~1, Ch.~4).
Upon maturation (only a few days under lab conditions) the zygospore germinates and gives rise to four vegetative cells by meiosis. These can be separated manually using a standard cell biology technique called tetrad dissection, and the subsequent progeny from the asexual reproduction of each of the four cells can be picked and cultured independently. This is a powerful and quite straightforward technique to combine genes from different strains. 

{\it Plus} and {\it minus} cells can only mate after becoming competent for sexual reproduction. 
The process is called  gametogenesis, and in the lab it can be induced easily by moving the cells to a medium without N-sources. The cells will then generally undergo a final round of cell division and then become gametes: they swim well and do not grow and divide. In principle, this produces a population of cells with very uniform properties, and in fact it has been used in several occasions in the past to study CR motility. Gametes' properties should remain constant for several days, but eventually they will die unless they mate or N is added back to the medium (in which case they revert to the vegetative state).

Culturing CR is reasonably straightforward. The cells grow easily at a temperature of $20^{\circ}-24^{\circ}$C and under ``white'' illumination of $\sim100\,\mu$E/m$^2$s (1E = 1 mole of photons). The medium can be completely inorganic, forcing phototrophic growth (i.e. cells need to photosynthesise to survive) or it can have acetate as organic carbon source. The latter is called myxotrophic growth and it is much faster (larger $n$ on average), but cell synchronisation is a bit poor.
There is a large CR community online, with many protocols readily available. A good place to start is ``The Chlamydomonas Resource Center'' ({\tt chlamycollection.org}), which contains also media recipes. Other useful algal collections include UTEX ({\tt utex.org}), SAG ({\tt uni-goettingen.de}) and CCAP ({\tt ccap.ac.uk}).

\section{The flagellar apparatus}
\label{sec:2}

\begin{figure}[t] 
   \centering
   \resizebox{0.8\columnwidth}{!}{
   \includegraphics{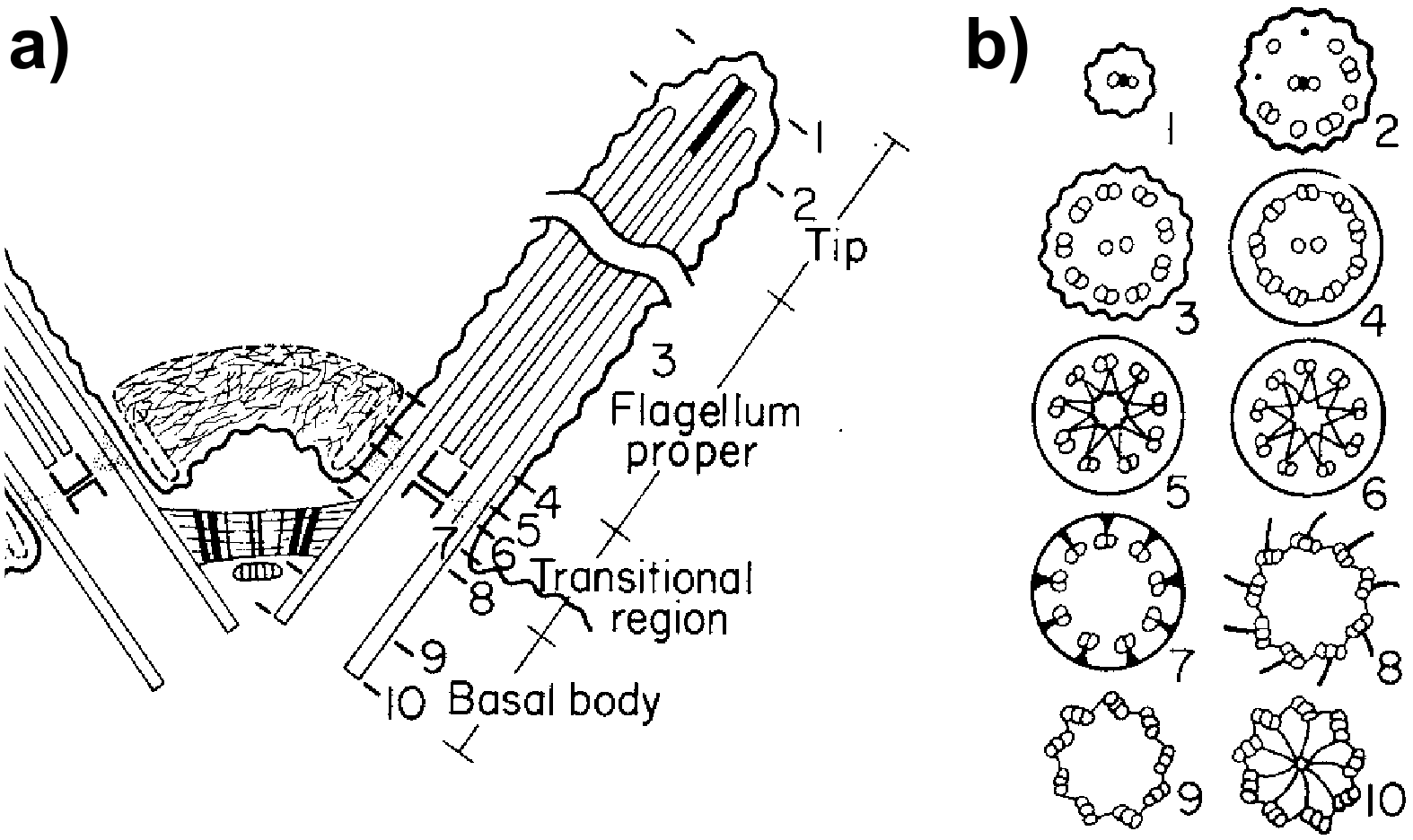}} 
   \caption{{\it Chlamydomonas} flagellar schematics 1. a) Spatial organisation of basal bodies, transition region and axoneme proper. Notice the drawing of the striated fibers connecting the basal bodies. b) Serie of sections of the flagellum at positions indicated by the numbers in a). (a,b) from \cite{harris09})}
   \label{fig2a}
\end{figure}

The two front flagella are arguably the most evident organelles in {\it Chlamydomonas} (see Fig.~\ref{fig2a}a). These are remarkably complex structures, composed by more than 500 different {\it types} of proteins \cite{pazour05}.  Flagella are used not just for motility but also for sensing (chemical and mechanical) and  mating. An astounding amount of work has poured into understanding CR flagella both at the molecular and dynamical levels, especially after it was discovered that much has been conserved throughout evolution and as a result several  human ciliopathies can be studied successfully in CR \cite{vincensini11}. Reviewing such an impressive body of work is well beyond the scope of these notes, which will necessarily only scratch the surface. A lot of what's beneath this surface has already been discovered... but much more is yet to be understood!

{\it Chlamydomonas} flagella are slender, whip-like objects $\sim12\,\mu$m long and $\sim0.25\,\mu$m thick, which exit the cell wall through specialised regions known as ``flagellar collars'' (see \cite{harris09} Vol.~1, Ch.~2). They are completely enclosed by the flagellar membrane, a domain of the cell's plasma membrane whose composition is highly regulated also by a diffusion barrier at the flagellar base (possibly realised by two structures known as the flagellar ``necklace'' and ``bracelet''). This membrane domain 
contains, among others, several types of voltage-gated and mechano-sensitive ion channels involved in phototaxis and perception of mechanical stimuli. Approximately $16\,$nm outside the flagellar membrane there is an extra ``fuzzy'' layer termed glycocalyx, a seemingly compact layer of carbohydrates connected to the most abundant flagellar membrane protein (FMG-1B). Two opposite rows of $0.9\,\mu$m-long, $16\,$nm-thick flexible filaments spaced $20\,$nm apart, protrude from the distal 2/3 of the flagellum. These so-called mastigonemes are exclusively found on the flagella of protists, where they are thought to increase hydrodynamic drag. 
It is not clear whether mastigonemes are  anchored just to the flagellar membrane or directly to the internal scaffold.
They are completely replaced every  $\sim4\,$hr. This is also the estimated turnover time for the flagellar membrane, which is continuously shed from the flagellar tip through  ectosomes possibly involved in cell-cell signalling \cite{wood13}. 

\begin{figure}[t] 
   \centering
   \resizebox{0.8\columnwidth}{!}{
   \includegraphics{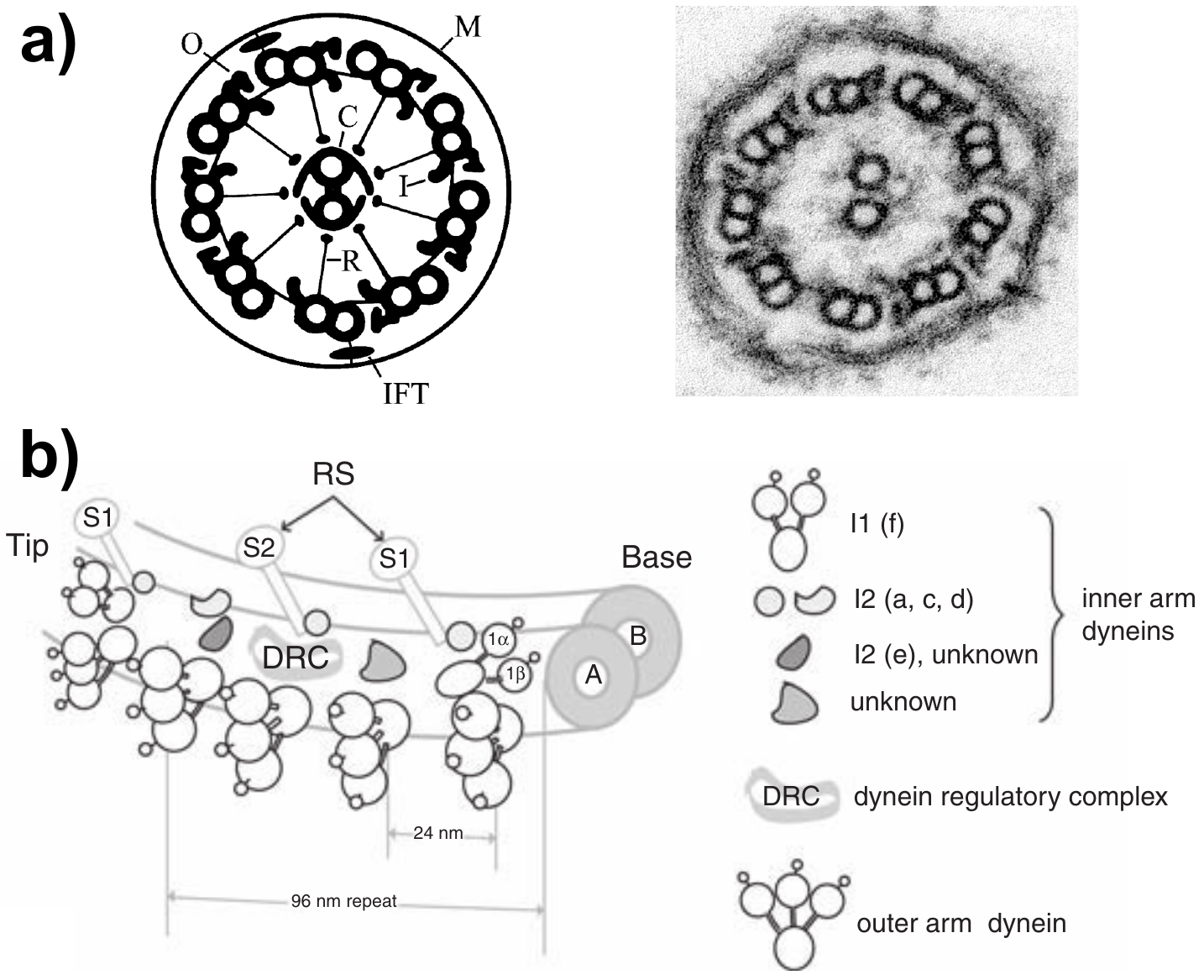}} 
   \caption{{\it Chlamydomonas} flagellar schematics 2. a) Schematics and electron micrograph of axonemal cross section. $O$: outer dynein arms; $I$: inner dynein arms; $C$: central pair; $R$: radial spokes; $M$: flagellar membrane; $IFT$: IFT trains. Notice the microtubule doublet (\# 1, doublet at the bottom of the schematic) without the outer dynein arm. b) Closeup of the basic $96\,$nm repeat unit on the outer doublets. $RS$: radial spokes ($S1$ and $S2$). (a) from \cite{pazour05}; b) from \cite{harris09})}
   \label{fig2b}
\end{figure}

The core structure within the flagellum is the axoneme, which is based on the standard 9+2 configuration: 9 microtubule doublets surrounding a central pair. The doublets, composed of microtubules A and B (13 and 11 protofilaments respectively) stem directly from the basal body triplets  mentioned earlier, through a characteristic transition region (Fig.~\ref{fig2a}a) displaying a ``stellate'' structure (sections 5,6 in Fig.~\ref{fig2a}a) typical of algae and sperm cells of land plants (e.g. ferns), but absent in protozoa or animals. Along the flagellum proper, the doublets are linked by nexins, $\sim40\,\mu$m-long polymers of a currently unknown protein. Notice that nexins are longer than the shortest distance between adjacent doublets ($\sim30\,\mu$m in straight axonemes). The central pair of microtubules is not connected to the basal body. It nucleates within the transition region  just above the stellate structure, which is thought to prevent the pair from sliding into the basal body itself. The two microtubules of the central pair are held together by bridges, and host a variety of proteins likely involved in flagellar metabolism, as well as kinesin motor proteins of unknown function. 
Along the portion of the central pair within the flagellum proper (see Fig.~\ref{fig2a}a), projections emerge at regular intervals based on (multiples of) a basic unit length of $16\,$nm. The projections interact with the head processes of the radial spokes, $30\,$nm-long rod-like structures attached to the A-tubules and extending towards the central pair. There is  strong evidence that this interaction is a key component in regulating the generation of bending moments within the axoneme (see below). 
The central pair is spontaneously twisted, a characteristic \underline{not} shared by animals, making a left-handed helix with $\sim2$ full turns along its length, and rotates during flagellar motion apparently being driven by bend propagation along the axoneme \cite{mitchell04}.
All axonemal microtubules are identically oriented with their plus-end towards the distal portion of the flagellum. While the central pair terminates precisely at the flagellar tip, the outer microtubules end $\sim0.5\,\mu$m earlier, first with the B and then the A tubules (sections 1,2 in Fig.~\ref{fig2a}a). The central pair and the A-microtubules terminate with different capping structures, all of which include plugs entering directly into the microtubule's lumen. The caps' function(s) is still unclear. Altogether, these passive components make for a rather stiff axoneme, with an estimated bending rigidity $\kappa\simeq 4\times10^{-22}\,$N\,m$^2$ \cite{niedermayer08}, which translates to a persistence length $\kappa/k_BT\sim10^5\,\mu$m ($\sim10^4\times$ flagellar length). 
The bending moments leading to flagellar motion are generated by axonemal dyneins, which localise on the A-microtubules and extend towards the B-microtubule of the nearest doublet (Fig.~\ref{fig2b}a). They are organised in two rows, the outer and inner dynein arms (``oda'' and ``ida''), depending on their position along the radius of the axoneme. Oda's and ida's are structurally different dyneins.
They are organised following the basic axonemal $96\,$nm repeat unit (Fig.~\ref{fig2b}b). This comprises 4 oda's, $24\,$nm apart, in the outer section; while in the inner section we have: a variety of ida's not yet completely characterised; two radial spokes; one dynein regulatory complex (DRC) which localises at the base radial spoke 2, plus several other regulatory proteins (mainly protein kinases and phosphatases). Linker proteins provide a direct physical connection between neighbouring oda's, between oda's and ida's, and between oda's and the DRC.  
One of the outer doublets lacks oda's: this is doublet 1 (Fig.~\ref{fig2b}a). The other doublets are numbered following the direction in which the dyneins extend. CR flagella are oriented with their doublets \#1 facing each other, and beat almost exactly along a plane determined by doublets 1, 5, and 6.

The process leading to microtubule bending is reasonably well understood \cite{riedel-kruse07b}: dyneins bridge between neighbouring doublets and use ATP hydrolysis to generate an inter-doublet sliding force which is converted to bending by the presence of geometric constraints to relative sliding of doublets (nexin links and basal body are the main suspects here). However, we currently do not have a clear understanding of either the basic mechanism leading to active oscillations (i.e. how does the system alternates the bending direction) or how such basic oscillation is then refined to give the observed waveforms. A solid body of experimental evidence shows that in {\it Chlamydomonas} the latter is achieved through active regulation of dynein activity by at least DRC and ida I1, which in turn are regulated by phosphorylation/dephosphorylation under the control of radial spokes/central pair. This active regulation adds a level of complexity which will be challenging to model, especially given that only indirect experimental data are available. In this context, high quality experimental characterisation of flagellar dynamics in CR will certainly be very useful. Experimental investigations based on analysis of flagellar dynamics in CR date back at least to the mid  80's \cite{ruffer85}, and have been recently refined by (semi)automated methods \cite{bayly10,bayly11,kurtuldu12,wan14}. They have provided tests for microhydrodynamics of cell locomotion (slender body vs. resistive force theory), achieved direct measurements of bend propagation along the axoneme, and characterised differences between wild type, {\it ida} and {\it oda} mutants. 
Comparing {\it wt} and mutants' flagellar dynamics and swimming behaviour  \cite{yagi05} revealed that ida's and oda's contribute differently to flagellar beating. Roughly, oda's are the workhorses of the flagellum, providing most of the internal power, while ida's are mostly responsible for the establishment of the correct waveform. As a consequence, {\it oda} mutants mostly have an altered waveform but close to normal beating frequency ($\sim50\,$Hz), while {\it ida} mutants have close to normal waveform but altered beating frequency ($\sim20\,$Hz). Clearly the separation is not completely clear cut. Still, despite these studies, the basic mechanism leading to flagellar oscillations eludes us. Currently, three alternative hypotheses have been put forward \cite{riedel-kruse07b,bayly15}: geometric clutch (GC), curvature control, and sliding control. These differ in the way dynein activity is periodically inhibited on opposite sides of the axoneme, a necessary condition for the emergence of oscillations. 
Ultrastructural analysis of quickly-frozen beating flagella in CR \cite{lindemann07} shows that when axonemes bend their cross section along the bending plane expands by $\sim25\%$. This observation would support the geometric clutch idea of oscillations caused by dyneins detachment induced by diverging transversal stresses within the axoneme. More recently the analysis of the most unstable beating modes in the three models also supported GC as a plausible basic mechanism for flagellar oscillation \cite{bayly15}. However, despite some evidence in support for GC, no definitive consensus has emerged yet. 

\section{Flagellar dynamics not related to beating}
\label{sec:3}

\begin{figure}[t]
\centering
\resizebox{0.75\columnwidth}{!}{
\includegraphics{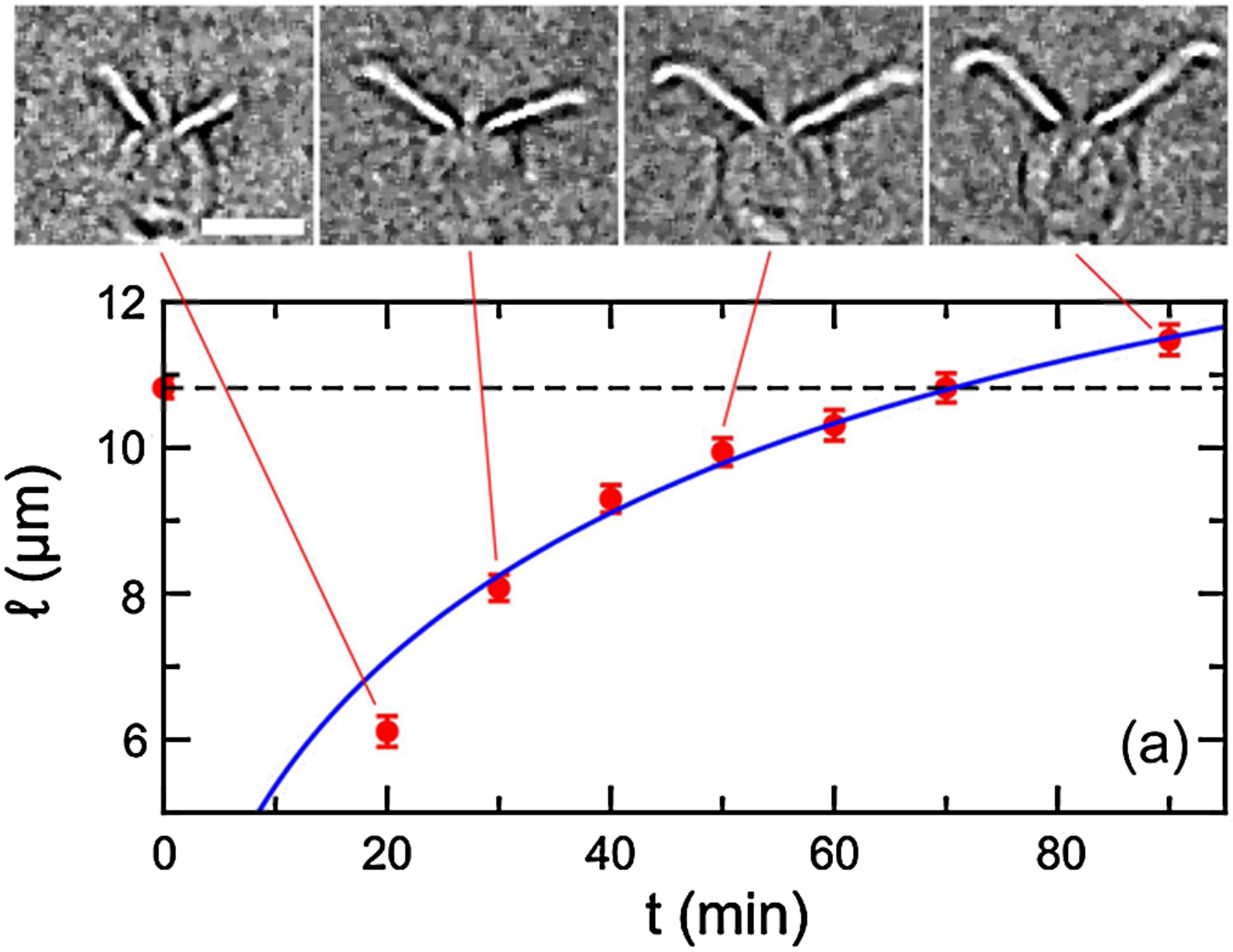}}
\caption{Flagellar elongation dynamics. Solid red circles: time evolution of flagellar length $\ell$ after mechanical deflagellation by pipette aspiration. Solid blue line: fit to the balance point model. The images of one of the recorded cells at different stages of regrowth have been processed to enhance contrast. Scale bar $10\,\mu$m. (From \cite{goldstein11})}
\label{fig3a}
\end{figure}

Looking under the microscope at a drop of {\it Chlamydomonas} culture deposited on a coverslip, it is common to see at the bottom surface many cells with their flagella spread wide apart and not beating. What is perhaps a bit more unexpected is that these cells, whose flagella adhere to the coverslip, move: this movement is called gliding. During gliding the cells slide at $\sim1.5\,\mu$m/s \cite{shih13} with the leading flagellum in front determining the direction of motion, and the other one trailing behind. The movement typically stops after a few seconds and when it resumes both flagella have the same probability to be the new leader. What drives gliding? 

The mechanism leading to this very peculiar kind of movement, which might have evolved before the actual axonemal beating \cite{shih13}, has only recently been demonstrated \cite{shih13,collingridge13} and -surprisingly- it is related to a seemingly completely disconnected phenomenon: the growth and maintenance of eukaryotic flagella.
We mentioned before that basal bodies, which connect flagella to the cell body, double up as centrioles during cell division. They cannot, however, perform both tasks simultaneously. In order to take part in cell division, basal bodies need to lose their flagella. This happens by an active resorption process, whereby the two flagella shrink simultaneously at a constant speed of $\sim0.1\,\mu$m/min, requiring a little over 2~hr to resorb fully grown flagella. Daughter cells then regrow their flagella before hatching, following a nonlinear growth dynamics (see below) that is completed over the course of $\sim3\,$hr. This dynamics can be studied very easily in {\it Chlamydomonas}. Pioneered in the late 60's by D. L. Ringo and J. L. Rosenbaum \cite{rosenbaum69}, studies of flagellar regeneration have relied on the fact that CR generally responds to a variety of harsh stimuli (including shear stress and pH shock) by shedding its flagella. This process, commonly known as flagellar autotomy or abscission, is induced by a Ca$^{2+}$ influx at the flagellar base \cite{wheeler07}. Calcium activates the microtubule severing protein katanin which cuts the axoneme at the ``site of flagellar autotomy'', a specific location within the transition zone. The evolutionary advantage conferred by the ability to actively cut the axoneme is not yet clear, but it certainly represent a big advantage for the experimental investigation of the dynamics of flagellar growth (as well as for proteomics of flagella).  
Fig.~\ref{fig3a} shows the typical regrowth dynamics. This has been studied mainly in paralysed flagella ({\it pf}) mutants, but wild type strains follow the same behaviour. During elongation the growth rate decreases monotonically, over the course of $2-3\,$hr,  from an initial value of $\sim0.4\,\mu$m/min to zero. Solid experimental evidence supports the idea that the growth rate depends on flagellar length, and not on time elapsed after the deflagellation.
During deflagellation by mechanical shearing, a small percentage of cells loses only one flagellum (``0-long'' cells). In this case the remaining one {\it shrinks} rapidly, up to $\sim0.4\,\mu$m/min, while the other regrows. Once they reach the same size, the flagella elongate symmetrically until the full length is recovered. This strategy is likely to have a direct impact on the swimming ability of these biflagellate cells, possibly minimising the time required to recover straight swimming, but this connection has not been studied. How do flagella grow? Why in the 0-long cells one flagellum shrinks while the other grows? Is there an active sensor of flagellar length? 

\begin{figure}[t]
\centering
\resizebox{0.99\columnwidth}{!}{
\includegraphics{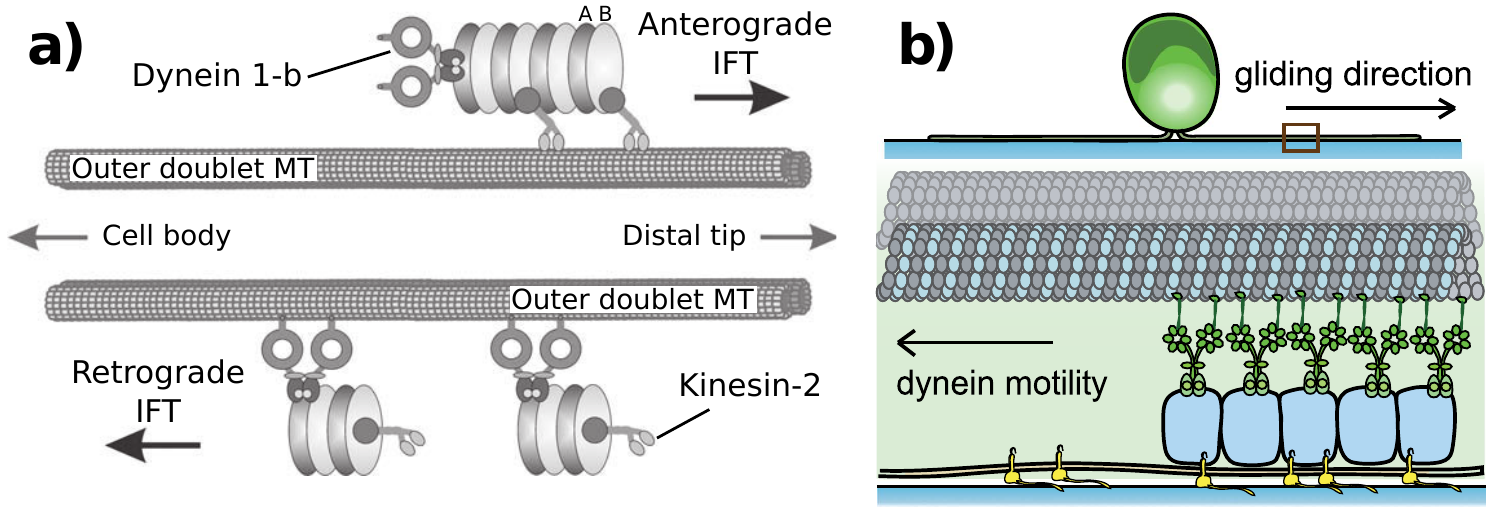}}
\caption{a) schematics of IFT trains' structure and motility. Anterograde IFT trains composed of stacks of A and B complexes are ferried towards the distal tip by kinesin-2 molecular motors. The IFT trains carry with them dynein 1-b motors which are responsible for retrograde motion towards the cell body. (Adapted from \cite{harris09}) b) Model for cellular gliding on a glass surface. The yellow transmembrane objects represent the protein FMG-1B. (From \cite{shih13}) }
\label{fig3b}
\end{figure}

The mechanism leading to flagellar growth has been shrouded in mystery until a serendipitous observation by K. G. Kozminski revealed a novel transport mechanism within the flagellum, aptly called the Intraflagellar Transport (IFT) \cite{kozminski93} (for an interesting historical account of the discovery see \cite{kozminski12}). IFT is highly conserved, and over the years it has proven to be, with few exceptions, the universal mechanism employed by eukaryotes to grow and maintain their flagella/cilia \cite{rosenbaum02,wemmer07}. As shown in Fig.~\ref{fig3b}a, it is composed of modular trains $0.05-1\,\mu$m long and $\sim50\,$nm wide \cite{pigino09}, walking incessantly along the outer microtubule doublets within the axoneme, just below the flagellar membrane. In fact, IFT trains were originally observed precisely because they make the flagellar membrane bulge out slightly. In wild type strains, IFT trains walk nearly always on the B tubule, rather than the A tubule where ida's and oda's are. Motion is both anterograde (towards the flagellar tip) with a typical speed $\sim 2\,\mu$m/s, and retrograde (towards the flagellar base) at a slightly higher speed $\sim 3\,\mu$m/s. 
The basic unit of the IFT trains is composed by one kinesin-2 and one cytoplasmic dynein-1b molecular motors, responsible for anterograde and retrograde motion respectively, which connect to the so called ``IFT complex'', composed of two parts, called A and B (themselves composed of several subunits \cite{harris09}. Not to be confused with A and B microtubules in microtubule doublets).
Within the transition zone, which acts as a diffusion barrier for cytoplasmic proteins of size larger than $\sim40\,$kD \cite{kee12} (but see also \cite{lin13}), axonemal proteins synthesised in the cytoplasm dock on specific sites on the IFT complexes within each train \cite{deane01}, and are then ferried all the way up to the flagellar tip. Some of these proteins, like dynein arms and radial spokes, already preassemble into complexes within the cytoplasm and only then are loaded onto IFT \cite{qin04}. At the tip the trains are remodelled, apparently breaking up into smaller units which then travel back to the base of the flagellum. The remodelling is associated with the release of the new axonemal proteins and the docking of ``turnover products'', old proteins that disassemble from the tips and are brought back to the cell body for recycling. IFT motility and protein shuttling do not stop when the flagellum reaches its full length: the axoneme converges to a dynamic equilibrium which needs to be maintained by a constant protein exchange between the flagellar tip and the cell body. In fact, if IFT is shut down, e.g. using temperature-sensitive mutants like {\it fla10}, full-length flagella will spontaneously disassemble at a constant rate of $\sim0.02\,\mu$m/min \cite{marshall05}, smaller than the case of ``active'' shrinking.
Even the set of proteins constituting the main part of the axoneme  are exchanged with new ones, at a rate of $\sim20\%$ every $6\,$hr. We now have direct experimental evidence that this happens by axonemal precursor proteins dissociating from IFT trains before reaching the tip, and then diffusing along the axoneme \cite{wren13}. The measured diffusivity ($\sim0.1\,\mu\textrm{m}^2$/s) is clearly significantly lower than what would be expected for a similarly sized particle in bulk water (a $2\,$nm radius sphere in bulk water at room temperature has a diffusivity $\sim100\,\mu$m$^2$/s).

This dynamic equilibrium has inspired a simple but quite successful model of flagellar growth: the balance point model \cite{marshall05}. Based on the discovery that the total amount of IFT proteins within growing flagella is {\it independent} of their length \cite{marshall01}, the model assumes that the number of IFT trains within a flagellum is constant, say $M$. A flagellum of length $L$ will have a growth rate $dL/dt = j_+-j_-$, given by the balance of an assembly current, $j_+$ and a dissociation current $j_-$. The latter is considered constant, as suggested by experiments. The former is given by $j_+ = p_{cargo}\, p_{int}\,\lambda/\tau$ where: $p_{cargo}$ is the loading probability of cytoplasmic precursor proteins onto IFT trains; $p_{int}$ is the assembly probability of released precursor proteins at the tip; $\lambda$ is a constant representing the length increase per new incorporated cargo; $\tau = 2L/Mv$ is the time between successive IFT train arrivals; and $v$ is the (harmonic) mean of anterograde and retrograde IFT velocities. Since $j_+$ decreases monotonically with $L$ while $j_-$ is fixed, the dynamics will have a single stable fixed point for  $L=L_*= Mv\lambda\,p_{cargo}\,p_{int}/2\,j_-$. 
Coupling $p_{cargo}$ to the size of the cytoplasmic pool of precursor proteins,
this model can successfully explain -at least qualitatively- flagellar growth dynamics in 0-long cells, dykarion rescue experiments, and mutants with variable flagella number ({\it vfl}) \cite{marshall05}. 
The balance point model has subsequently been revised \cite{engel09} following new experimental results which suggest that, although the total amount of IFT protein is independent of flagellar length, the number of IFT trains increases with $L$. This was interpreted as a remodulation of the average size of the IFT trains, but no explicit mathematical model has been put forward. However, recent experiments have questioned the validity of the balance point model altogether, proposing instead a process based on differential cargo-loading of IFT trains, possibly under direct control of a length sensor \cite{wren13}. Flagellar length is in fact well known to be {\it also} under genetic control \cite{tam13}, through the expression of several kinds of non-IFT proteins, mainly kinases and phosphatases, some of which localise in specific regions within the cytoplasm but whose {\it modus operandi} is at the moment completely unknown.

So how does IFT relate to gliding? It turns out that IFT trains are also coupled to proteins on the flagellar membrane, a connection which implies a major role for IFT  in flagella-mediated processes like mating \cite{harris09}. In particular, they associate with the FMG-1B membrane glycoprotein mentioned above (see Fig.~\ref{fig3b}b; this connection is usually transient, through some unknown linker protein which is Ca$^{2+}$ sensitive \cite{collingridge13}). As the flagella adhere to the surface, the sugar moieties of FMG-1B  can stick to glass preventing IFT trains from moving along the microtubules. These will then pull the axoneme in the direction opposite to their previous motion, similarly to what happens with focal adhesion points and acto-myosin cell motility in mammals.  
Only IFT dyneins seem to be involved in this process, so the force will pull the cell towards the distal tip (plus end) of the leading flagellum. 
Direct force measurement with optical tweezers holding colloids bound to FMG-1B, measured forces of $20-30\,$pN, suggesting that these adhesion points are clusters of $\sim4$ motors.
Gliding will then cease whenever the dyneins manage to detach from their constraint either by disassociating from FMG-1B or because they simply reach the base of the flagellum, where the IFT trains are recycled.

\section{Flagellar motility: Synchronisation and Swimming}
\label{sec:4}

Despite the importance of IFT, the most immediately striking type of dynamics displayed by flagella is certainly their incessant beating. {\it Chlamydomonas} flagella follow mostly a so-called ``ciliary'' type of beating, with bending waves which propagate along the axonemes \cite{bayly11,kurtuldu12} and cause the continuous alternation of well defined power and recovery strokes. A characteristic feature of this dynamics is the pronounced synchrony of the two flagella, which usually phase lock for seconds on end, although the exact average duration depends strongly on flagellar length \cite{goldstein11}.
 This is most often ``in-phase'' locking, but the phototaxis mutant {\it ptx1} was recently shown to display also extended periods of anti-phase synchronisation, which are however associated with a slightly different waveform \cite{leptos13}. It is not known whether there is a causal connection between the type of phase-locking and flagellar waveform.
 
What causes phase locking? In the last few years, this problem has stimulated quite a lot of work, both experimental and theoretical. 
Experimental investigations have been based mainly on long-time high-speed recordings of flagellar motion in pipette-held cells \cite{goldstein11,leptos13,goldstein09}. Following the lead of pioneering studies by U. R\"uffer and W. Nultsch in the mid 80's \cite{ruffer85}, these studies have revealed that normal phase synchrony is noisy, and that noise can occasionally lead to phase slips: brief lapses of synchrony lasting a few beats ($\lesssim100\,$ms) whereby one flagellum accumulates one or more full extra cycles with respect to the other. Either flagellum can slip ahead, although the probability is usually biased to a cell-dependent-degree towards a specific flagellum. These observations can be recapitulated very well using a simple effective model, where the flagellar phase difference $\Delta(t)$ evolves according to $\dot{\Delta}(t) = \delta\nu - 2\pi\epsilon\sin(2\pi\Delta) + \xi(t)$. Here $\delta\nu$ is the intrinsic frequency difference between the flagella (responsible for the slip bias), $\epsilon$ is their effective coupling, and $\xi$ is an effective noise term responsible for the slips. 
For $|\delta\nu|<2\pi\epsilon$ the system has two fixed points, one stable and one unstable, the stable one representing the observed state of phase locking.
This stochastic Adler equation can actually be derived as the first order description of the generic dynamics of weakly coupled self sustained phase oscillators \cite{pikovski}, so in a sense it is not completely surprising to find it here. However, different coupling mechanisms will produce $\epsilon$'s of different magnitude and which depend differently on parameters of the system, and so experiments that change $\epsilon$ can in principle be used to determine what is the origin of the observed coupling. Two main models have been proposed: 1) the coupling comes from the interplay between direct hydrodynamic interaction and elasticity intrinsic in the waveform \cite{niedermayer08}; 2) the coupling results from modulations of flagellar driving force within a beating cycle \cite{uchida11}. 
Although the relative strength of coupling from these two effects can be tuned within colloidal systems of rotors \cite{kotar13}, experimental tests with  somatic cells of the multicellular species {\it Volvox carteri} \cite{brumley14}, a relative of CR, support clearly an elasto-hydrodynamic origin for the synchronisation observed between flagella mounted on different cells. Until recently, this seemed to be the case also for the two flagella of a single {\it Chlamydomonas} cell \cite{goldstein11}, but new experiments point instead to a fundamental role played by the distal striated fibres connecting the basal bodies directly \cite{quaranta16,wan16}, opening an interesting chapter in our understanding of the roles played by mechanical forces within the cell.
An alternative model based on cell-body rocking has been proposed by B. Friedrich and coworkers \cite{geyer13}, and we refer the reader to his chapter for more informations.
Still, the observation of prolonged alternate periods of both in-phase and anti-phase synchronisation in {\it ptx1} poses new challenges to our understanding of flagellar synchronisation, currently not solved \cite{polin15}. The key to understanding the problem will come perhaps from experiments specifically characterising flagellar beating noise \cite{wan14}.
Besides normal flagellar movement, {\it Chlamydomonas} can also display a characteristic ``shock'' response, where the flagella undulate in front of the cell in a ``flagellar'' type motion (snake-like). This shock lasts $\sim500\,$ms, and is triggered by a massive Ca$^{2+}$ influx within the flagellum \cite{hegemann89} in response to intense stimuli. Interestingly, during shock dynamics flagella hop between periods of in-phase and anti-phase synchronisation, but this aspect has not been studied in detail yet. 

Hydrodynamic models of flagellar waveforms suggest a reason for evolving a separate shock response: this seems to be optimised for fast escape, while normal beating is optimised for feeding \cite{tam11}. However, during shock response the cell does slow down noticeably ($20\,\mu$m/s vs. $100\,\mu$m/s), so the connection with a more efficient escape is not immediately clear.
During normal flagellar dynamics, {\it Chlamydomonas} swims along a tight left-handed helix, caused by small chiral tilts in the waveforms of its flagella. The spinning frequency is $\sim2\,$Hz, with a resulting pitch of $\sim50\,\mu$m. For a detailed mathematical description of helical swimming see \cite{crenshaw89}. As the cell moves on a helix, its eyespot continuously scans the environment. Swimming within a light field, then, produces a temporally modulated signal whenever the cell is swimming at an angle different from $0$ or $\pi$ with respect to the direction of light propagation. This is the basic signal used for phototactic steering of the helical trajectory \cite{harris09}. However, even without external stimuli the helix is clearly not perfectly straight! Active flagellar noise (e.g. phase slips, but the actual origin has not been explicitly investigated experimentally) causes a small amount of angular diffusion $D_{rot}$, which has been measured explicitly for the close species {\it C. nivalis} to be $D_{rot}\simeq2\,$rad$^2$/s \cite{hill97}. By itself, this would cause a spatial diffusivity $D\simeq 0.25\times10^{-4}\,$cm$^2$/s \cite{sevilla15}. This is significantly smaller than the value $D\simeq 7\times10^{-4}\,$cm$^2$/s which has been measured directly on a population of {\it C. reinhardtii} \cite{polin09}. The discrepancy is due to the fact that the effective angular diffusion is not the main mechanism leading to CR spatial diffusion. Instead, diffusion is dominated by sharp reorientations which happen randomly following a Poissonian dynamics with characteristic time $\sim10\,$s. These sharp turns are due to $\sim2\,$s intervals during which CR flagella loose synchrony and beat at a constant but $\sim30\%$-different frequency. This is probably due to a substantial increase of their intrinsic frequency difference \cite{polin09} rather than a weakening of interflagellar coupling, possibly caused by changes in cytosolic Ca$^{2+}$ concentration. Although the origin of this phenomenon is not well understood, it provides a direct evidence that the cell is capable of actively modulating the synchronisation state of its flagella.

\section{Interaction with boundaries}
\label{sec:7}

\begin{figure}
\centering
\resizebox{0.95\columnwidth}{!}{
\includegraphics{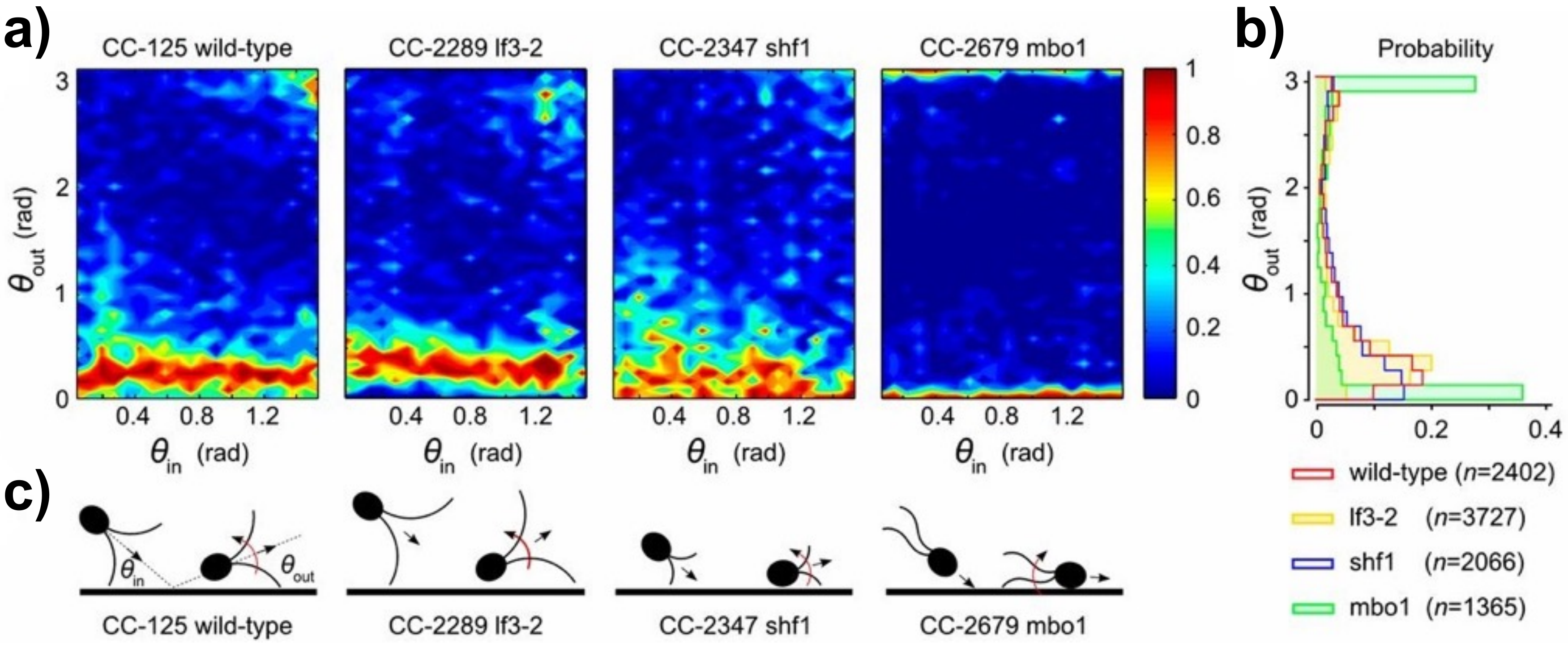}}
\caption{Dynamics of CR scattering off a planar wall. The incoming and outgoing scattering angles,  $(\theta_{in},\theta_{out})$, are calculated from the plane.  a) Conditional probability $p(\theta_{out} | \theta_{in})$ for different CR mutant strains in the flat wall experiment of \cite{Kantsler12}. b) Distributions of $\theta_{out}$ for all $\theta_{in}$. c) Schematic illustration of the flagella-induced scattering mechanism. The $mbo1$ (``moving-back-only'') mutant is trapped at the boundary for long times.}
\label{fig_scattering1}
\end{figure}

How do microorganisms interact with the physical surfaces that surround them? Can we conceptualise these interactions as essentially hydrodynamic, or essentially steric, or do we need a combination of both? Is this the same for all types of microorganisms? Can we use this knowledge e.g. to design surfaces that will be difficult for microorganisms to colonise?

Starting with the seminal 1963 observations by Rotschild \cite{Rotschild}, who reported on the accumulation of sperm cells on the sides of a channel, it is now a well established fact that pusher-type microswimmers (bacteria, sperm and other swimmers with rear-mounted flagella) accumulate on flat solid boundaries. Both purely steric \cite{Li09} and purely hydrodynamic \cite{Berke08,Li14} explanations are in good agreement with experimental data. Recent experiments, however, have finally demonstrated that wall accumulation of bacteria is essentially a hydrodynamic phenomenon  \cite{Sipos15} (but see also \cite{Molaei14}). Looking at swimming of {\it Escherichia coli} in presence of cylindrical obstacles similar to those we will discuss below, O. Sipos and collaborators \cite{Sipos15} demonstrated for the first time that, as predicted by hydrodynamic interactions, convex surfaces of a sufficiently small curvature trap bacteria by locking their swimming at an inward angle towards the surface. Neither of these phenomena could be explained by purely steric interactions.

\begin{figure}
\centering
\resizebox{0.9\columnwidth}{!}{
\includegraphics{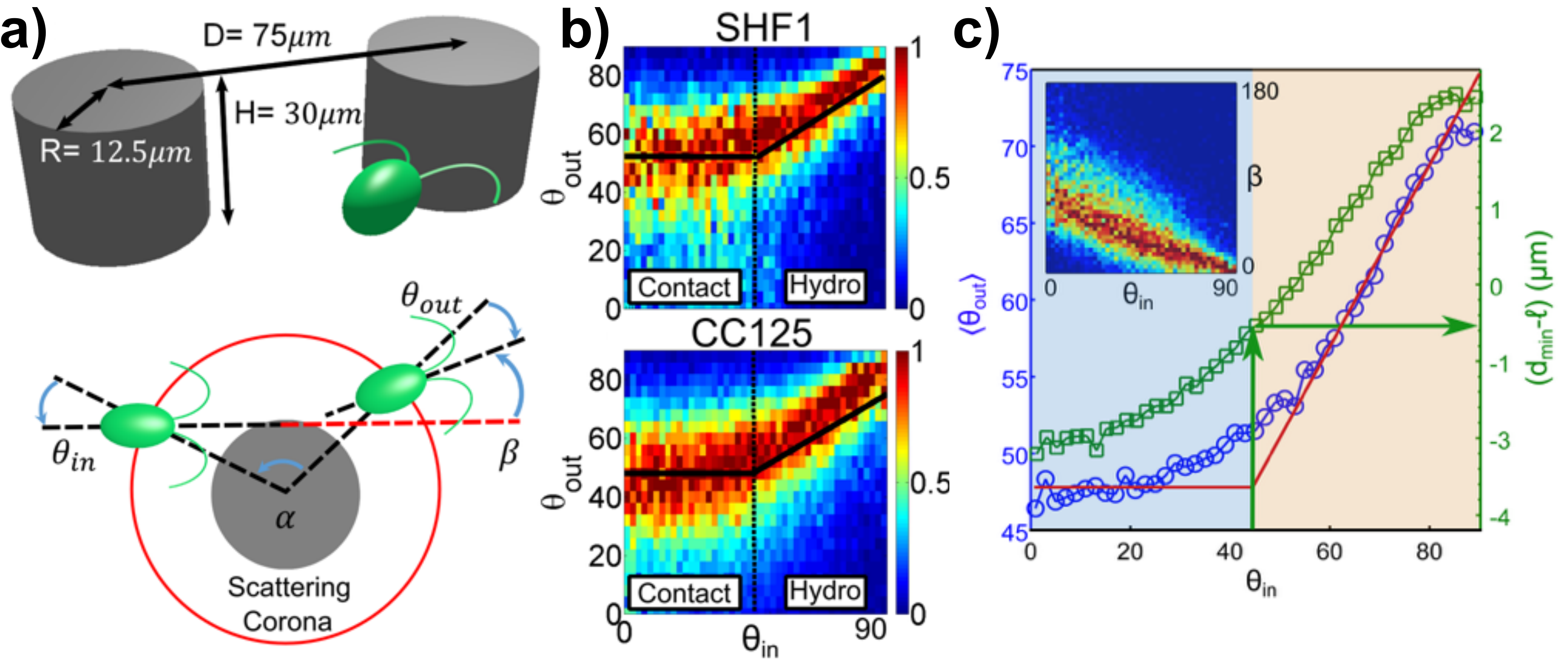}}
\caption{a) Schematic of the experimental setup used in \cite{Contino15} and definition of the scattering angles $(\theta_{in},\theta_{out})$. Notice that the scattering angles are defined differently from Fig.~\ref{fig_scattering2}. b) $p(\theta_{out}|\theta_{in})$ for wild type (CC125) and short flagella mutant (SHF1). The ranges of $\theta_{in}$ corresponding to hydrodynamic and contact regimes are highlighted. c) $\langle\theta_{out}\rangle$ (blue circles) and $\langle d_{min}-l\rangle$ (green squares) as a function of $\theta_{in}$. The inset shows the net angular deviation $\beta$ of the swimmer's trajectory as a function of $\theta_{in}$.}
\label{fig_scattering2}
\end{figure}

Swimmer-wall interaction for microorganisms with front-mounted flagella like {\it Chlamydomonas} (puller-type) is distinctly less understood. 
Analysing the scattering of different CR strains off a flat surface within a thin microfluidic channel, V. Kantsler \textit{et al.} \cite{Kantsler12} observed for the first time that their escape angle from the wall is constant and essentially independent of the angle at which they approach the surface (Fig.~\ref{fig_scattering1}a,b). Its value can be predicted with reasonable accuracy simply as the angle between the cell's longitudinal axis and the line joining the back of the cell body with the tip of the flagellum at its maximal extension (Fig.~\ref{fig_scattering1}c). This is the hallmark of a fundamentally steric interaction, dominated by direct flagellar contact with the surface.  
Recent experiments, however, show that this simple picture is not complete  \cite{Contino15}. Looking at CR scattering off cylindrical pillars within a microfluidic device, and monitoring the dependence of the outgoing angle $\theta_{out}$ on the incoming angle $\theta_{in}$, here measured from the local surface normal (Fig.~\ref{fig_scattering2}a), the authors observed two distinct types of interaction: a hydrodynamic regime for large $\theta_{in}$, and a contact one for small $\theta_{in}$ (Fig.~\ref{fig_scattering2}b). 
%
Within the hydrodynamic regime, $\theta_{out}=m\theta_{in}+q$ (Fig.~\ref{fig_scattering2}c), with $m\simeq0.6$. A value of $m\neq1$ signals the presence of an interaction, which is also revealed by a net deflection of the swimmer trajectory (Fig.~\ref{fig_scattering2}c inset). At the same time, the minimal distance of the swimmer from the surface $d_{min}$ is always larger than the CR's flagellar length, ruling out direct contact with the pillar. Within this regime, then, the microswimmer interacts with the obstacle only hydrodynamically. 
Conversely, within the contact regime $\theta_{out}$ is independent of $\theta_{in}$, a result identical to the flat boundary case \cite{Kantsler12}. However, looking at the process more carefully reveals that this type of scattering is complex, and includes a hydrodynamic contribution. 
During a typical scattering within the contact regime, the cell hits the pillar surface and is then reoriented along the local tangent direction with the flagellar plane parallel to the surface. Lubrication forces then keep the cell swimming close to the pillar until CR spinning rotates the flagellar plane by $90^{\circ}$. This orientation maximises flagellar push against the solid boundary, and the alga then leaves the obstacle through the same mechanism observed for flat surfaces. The whole process is a mixture hydrodynamic interactions, which would tend to trap the organism around sufficiently large obstacles, and escape by direct flagellar contact. Being able to avid long term trapping at surfaces might in fact represent an advantage for a soil dwelling microorganism like {\it Chlamydomonas}.

\section{Conclusion}
\label{sec:8}

We have presented a brief introduction to biophysical studies of {\it Chlamydomonas reinhardtii}, one of the main model organisms in biology. Clearly, we have literally only scratched the surface, and did not mention many phenomena which are at least as interesting and important as those presented. At the single cell level, most notably we did not talk about modulations of swimming by either active response to stimuli (phototaxis, chemotaxis) or passive (gravitaxis and gyrotaxis). These can in turn induce interesting phenomena at the population level, like the emergence of bioconvective patterns. We do hope, however, to have stimulated the readers' curiosity to know more about this fascinating microscopic organism. It certainly has still a lot to offer to the careful and interested researcher.

\section*{Acknowledgements}
The final publication is available at Springer via http://dx.doi.org/XXXX
%

\end{document}